\documentclass[preprint]{aastex7}
\usepackage{graphicx} % Required for inserting images
\usepackage{natbib}
\usepackage{xcolor}
\begin{document}
\title{Two Be or Not Two Be\footnote{``To Be or not to Be" was first used as a pun by \citet{Bidelman1976} in during his closing remarks at IAUS No. 70}: A New Companion detection for HD 52244 using HST/FGS}

\author{Keefe J. Kamp}
\affil{Embry-Riddle Aeronautical University}
\email{keefe.kamp@my.erau.edu}

\author{Saida M. Caballero-Nieves}
\affil{Embry-Riddle Aeronautical University}
\email{Saida.Caballero@my.erau.edu}

\author{Edmund P. Nelan}
\affil{Space Telescope Science Institute}
\email{nelan@stsci.edu}

\author{Nancy Remage Evans}
\affil{Harvard Smithsonian Center for Astrophysics}
\email{nevans@cfa.harvard.edu}

\author{Douglas R. Gies}
\affil{Georgia State University}
\email{dgies@gsu.edu}

\author{Noel D Richardson}
\affil{Embry-Riddle Aeronautical University}
\email{RICHAN11@erau.edu}

\begin{abstract}
In this paper we present a newly detected companion to the Be star, HD 52244 (B2IVnpe), using the Fine Guidance Sensors (FGSs) on the Hubble Space Telescope (HST). In fall 2021, HST became momentarily unavailable to support nominal operations, and we used the operational FGS to carry out a multiplicity survey of 6 Be stars. We were able to resolve a companion to HD 52244, with a separation of 42.7 ± 1.1 mas (74 AU) and a position angle of 144.2 ± 0.3 with a differential magnitude in the F583W filter of 1.91 ± 0.02 mag. This study presents the results to the newly detected companion of HD 52244 and lays the groundwork for future studies looking for wide or third companions to Be stars. 
 
\end{abstract}
\section{Introduction}
There exists a unique subset of B type stars called classical Be stars. These stars are defined by the presence of hydrogen emission lines in their spectra, predominately H$\alpha$. According to current consensus, this emission is due to a circumstellar decretion disk caused by near critical rotation of the B star. One way to spin up a B star is via the exchange of angular momentum from tidal interactions with a binary companion \citep{Rivinius2013}. Thus, the study of Be star binarity can provide important information on the role of binaries. In addition, \citet{Tokovinin1997} found that short period binaries were often in a triple system resulting in angular momentum being placed into the wide orbit of a third companion. However, B stars, let alone Be stars, have been neglected when it comes to probing their multiplicity. Their intrinsic variability \citep{Gautschy1993} makes it hard to disentangle binary signatures in both their light curves and spectra. These stars are also dimmer than O stars, and less populous than the lower mass G and M stars, further making it harder to probe their multiplicity. Binaries are an essential byproduct of the star formation process \citep{Zinnecker2007}. More than 70\% of the most massive stars are known to have at least one companion \citep{Mason1998}. This fraction decreases to about 50\% for sun-like G stars \citep{Raghavan2010} and all the way down to 10.3\% for M dwarfs. \citet{Dunstall2015} looked into the multiplicity of B stars in the Large Magellanic Cloud using spectroscopy. Their study found the binary fraction of this population of B stars to be 25\% for companions with periods less than $10^{3.5}$ days and mass ratios greater than 0.1. When they took observational biases into account, using synthetic populations, they estimated the multiplicity fraction to be 0.58. For a Galactic population \citet{Abt1990}, looked at main sequence B stars from B2-B5 resulting in a bias corrected binary fraction of 0.74 taking into account distances covered by spectroscopic to visual binaries. When it comes to Be stars, many studies (\citealt{Dulaney2017}; \citealt{Peters2016}; \citealt{Peters2013}; \citealt{Abt1984}) have searched for companions. However, these studies have been limited in scope and were all spectroscopic surveys. One study by \citet{Mason1997} used speckle interferometry in a similar scope to this investigation and found a binary fraction of 10\%.  \citet{Bodensteiner2019} cross-matched 287 Be stars in literature and determined a 8.36\% of the stars have some form of companion, none of which are main sequence. These multiplicity fractions are much lower than we expect from the trend we see from O to G to M, decreasing multiplicity with decreasing mass. This would suggest that the multiplicity of Be, and subsequently B stars, need to be further investigated. In this study we use a high angular resolution method in order to probe shorter periods than we could for the galactic O stars at the same angular resolution as O stars are farther. Thus allowing us to start filling in the gap in their period distribution for massive stars.

This study is a high spatial resolution survey of six Be stars using the Hubble Space Telescope (HST) Fine Guidance Sensors (FGS). In section \ref{sec:Sample} we will describe our sample of 6 Be stars as well as the observations with HST/FGS in section \ref{sec:Obs}. In section \ref{sec:Comp Dect} we go into the methods used to detect the binary companion. Next in section \ref{sec:Discussion} we discuss the multiplicity of the sample, mainly the detection of a companion of HD 52244. Finally in section \ref{sec:sum+FW} summarize our findings and discuses future works for the multiplicity of Be stars.

\section{Sample of Be Stars}
\label{sec:Sample}
\begin{figure*}
    \includegraphics[width=0.75\textwidth]{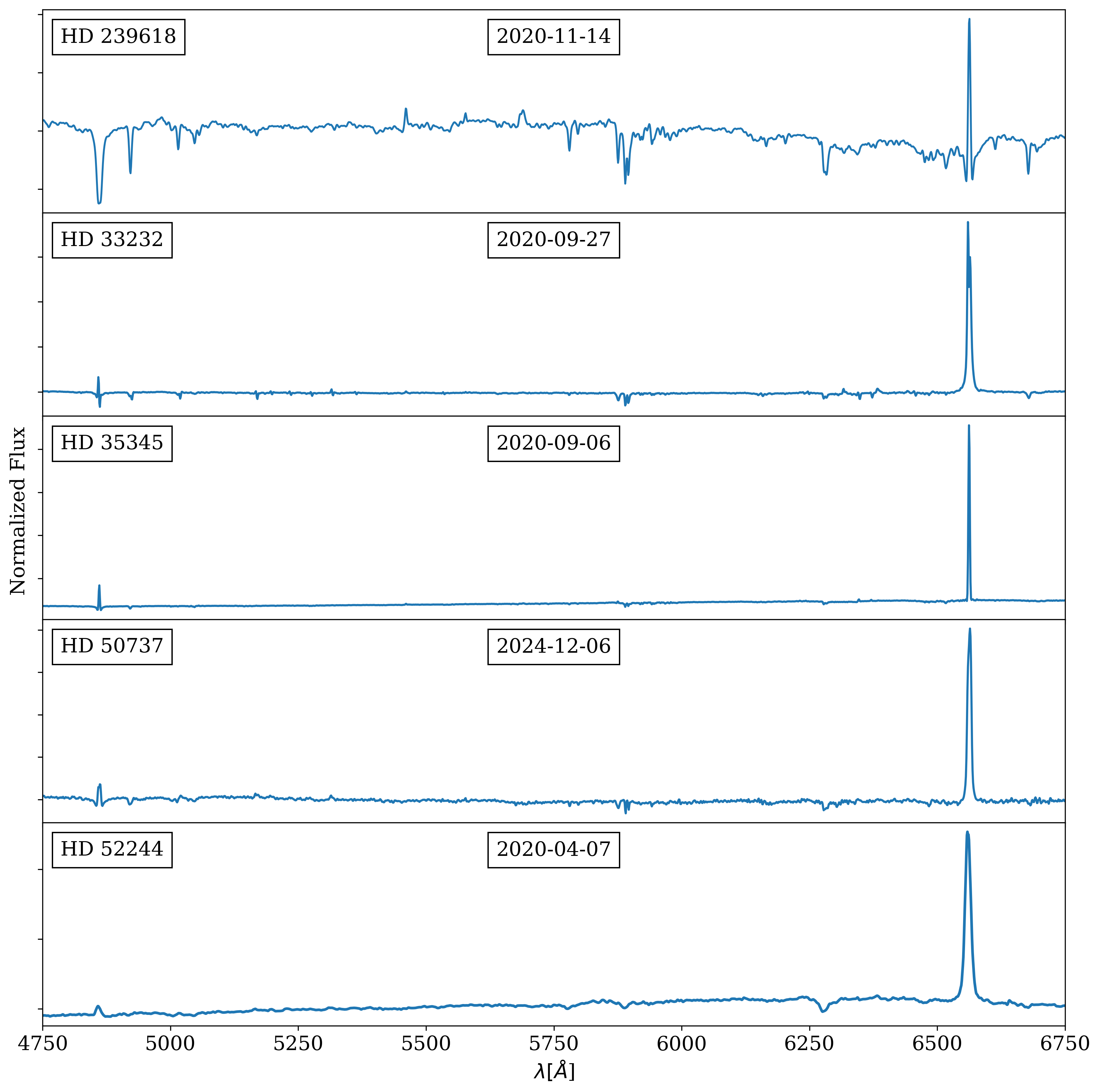}
    \includegraphics[width=.14\textwidth]{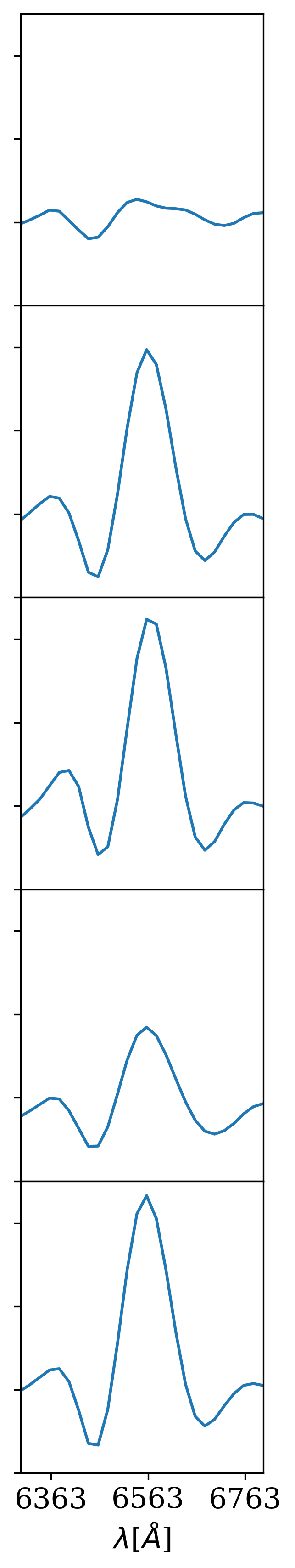}\\
        \includegraphics[width=0.75\textwidth]{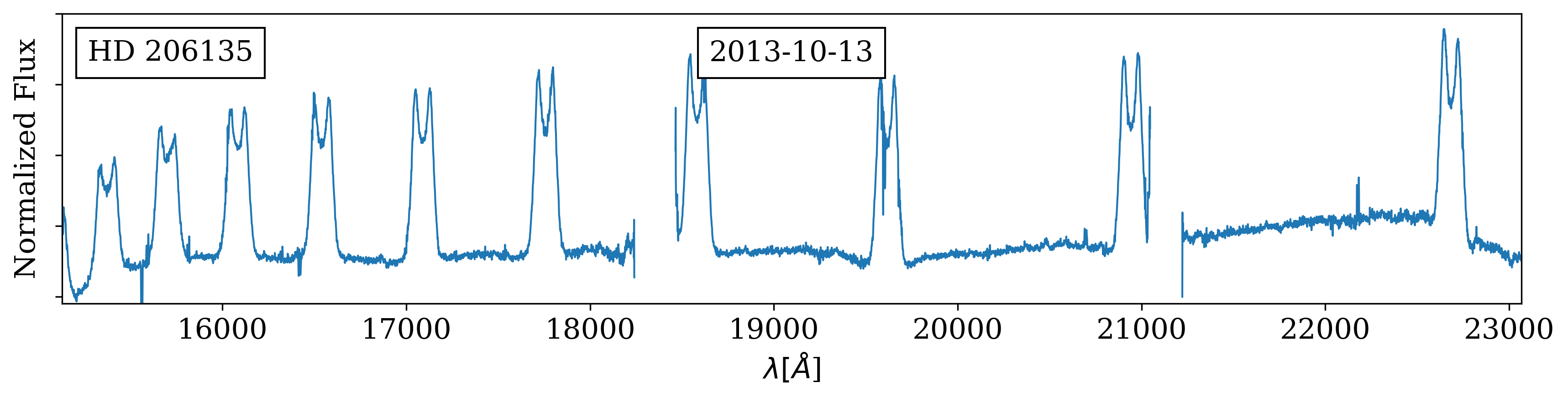}
    \includegraphics[width=.14\textwidth]{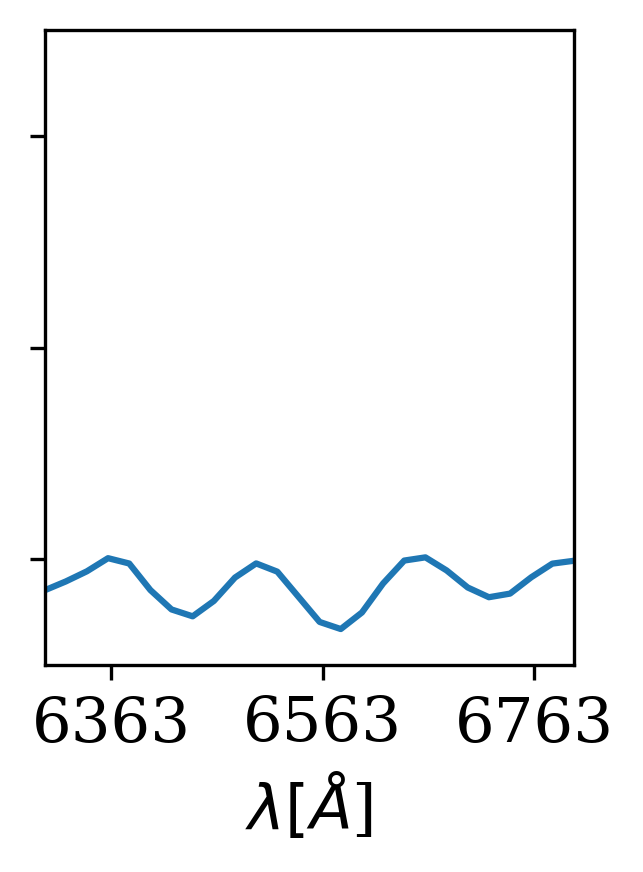}
    \caption{Presented here are the spectra of the six Be stars in this study. To the left is the visible light spectrum (with the exception of HD 206135 where it is IR from APOGEE ;\citealt{Apogee}) of the 6 of Be from the BeSS Database \citep{Neiner2018} and the right postage stamp size plots are around the H$\alpha$ emission from the low resolution spectra from \citet{GaiaDR3}. \label{fig:spec}}
\end{figure*}

We performed high spatial resolution observations of six Be stars using HST/FGS during Cycle 29 (HST proposal \#16868 PI: S. Caballero-Nieves) during the Science Instrument Command and Data Handling module failure. These targets are tabulated in Table \ref{tab:Targets}.  In Figure \ref{fig:spec}, we show the spectra of all six stars using the available BeSS \citep{Neiner2018} visible echelle spectra with two exceptions: HD 52244 which is a non-echelle visible spectra; and HD 206135 where it is an infrared Spectra from SDSS/APOGEE \citep{Apogee}. We selected these spectra based off of their proximity to the date of observations with FGS. The postage stamp sized plots to the right are from Gaia low dispersion spectra cropped around the H$\alpha$ emission line ($\lambda_{vac} \approx 6562$\AA) in order to highlight any H$\alpha$ emission usually used to identify Be stars. The top four stars show the characteristic H$\alpha$ line in emission. In the bottom two it is less obvious, however, there are still classified as Be stars. HD  239618 has a small emission bump present in the Gaia postage stamp, however, in the full spectrum to the left (taken at a presumably different date) there is an emission line present at  6562\AA. When it comes to HD 206135 there does not appear to be H$\alpha$ emission present in the Gaia spectrum, however, \citet{Chojnowski2015} classifies it using Br11 in the infrared  ($\lambda_{vac}=16811$\AA). This line is the strongest hydrogen emission in the infrared and not affected by afterglow or telluric effects \citep{Chojnowski2015}. The double peaked shape of the emission lines present in this spectra further support the emission line coming from a circumstellar decretion disk \citep{Chojnowski2015}. This phenomena is sometimes referred to as a ``shell star'' however this is now believed to be the same phenomenon as Be stars \citep{Rivinius2013}.

Table \ref{tab:Targets} provides the following details for the six observed targets from left to right are: the HD number identifier of the target; the right ascension and declination of the targets in J2000; the spectral class of the target; the reference to the spectral class; the V magnitude of the target; the B-V color of the target; the parallax from Gaia DR3; the RUWE from Gaia DR3; the NEOWISE W1-W3 color; the date the observation; the name of the primary calibrator used, in the case of HD 52244 (discussed further in the next section).

\begin{deluxetable}{ccclccrlcccc}
\label{tab:Targets}
\tablenum{1}
\tablecaption{Be Star Target List}
\tablewidth{0pt}
\tablehead{
\colhead{HD} & \colhead{RA} & \colhead{Dec} & \colhead{Spectral} & \colhead{Class} &\colhead{V} & \colhead{B-V} & \colhead{Gaia Plx.}&\colhead{RUWE}&\colhead{W1-W3}&\colhead{Obs. Date} & \colhead{Calibrator} \\
\colhead{Number} & \colhead{(J2000)} & \colhead{(J2000)} & \colhead{Class.} & \colhead{Ref} &\colhead{(mag)} & \colhead{(mag)} & \colhead{mas}&\nocolhead{}&\colhead{(mag)}&\colhead{YYYY-MM-DD} & \colhead{Name}
}
\decimalcolnumbers
\startdata
33232 & 05 10 48.2024 & +41 00 10.431 & B2Vne & B & 8.16 & 0.07 & 0.7552&0.86&2.00 &2021-11-16 & \nodata \\
35345 & 05 25 44.7841 & +35 38 49.911 & B1Vpe & C & 8.43 & 0.11 & 0.8841 &0.96&1.60&2021-11-19 & \nodata  \\
50737 & 06 53 52.1678 & -13 11 09.112 & B2Vnne & A & 8.68 & -0.04 & 0.7942 &1.02&1.51 & 2021-11-21 & \nodata  \\
52244 & 06 59 46.4591 & -16 12 02.728 & B2IVnpe & A & 9.68 & -0.06 & 0.5764&1.84&1.15 &2021-11-20 & HD  50737   \\
239618 & 21 14 45.4840 & +59 45 39.707 & B2Ve & D & 8.65 & 0.4 & 1.0129&1.85&1.20 &2021-11-22 & \nodata  \\
206135 & 21 36 57.0541 & +68 11 07.301 & B3V & E & 8.27 & 0.12 & 1.0987&0.92&0.40&2021-11-16 & \nodata \\
\enddata
\tablecomments{(A) \cite{Levenhagen2006} (B) \cite{Guetter1968} (C) \cite{Morgan1953} (D) \cite{Boulon1959} (E) \cite{Racine1968}}
\end{deluxetable}
\newpage
\section{Observations}
\label{sec:Obs}
These observations were approved as director's discretionary proposal to fill otherwise unused telescope time while STScI switched the  HST's Science Instrument Command and Data Handling (SI C\&DH) module (which operates the HST's science instruments) to its ``side-B" configuration. During this time only the Fine Guidance Sensor (FGS) could be used for scientific observations. The FGS is a dual-axis white light, broad band shearing interferometer that utilizes a K{\"o}ster's prism, that when used in its Transfer Mode, can detect modest contrast binary systems with angular separations down to $\sim10$ milli-arcseconds. Detailed description of the FGS as a science  instrument can be found in \citep{InstHandbook}.  For this proposal we did not want to  command a mechanism move, i.e., a filter wheel move, given that the FGSs are aging mission critical instruments. This lead us to select FGS2 for our science  observations as this has in place the F583W full aperture filter element used for guiding, which also offers the best throughput and angular resolution for our observations. However, this also restricted us to select Be stars fainter than $\sim8$th magnitude to avoid saturating the instrument's photomultiplier tubes. Observing brighter stars would have required commanding the filter wheel to place the F5ND attenuator in the beam. This constraint yielded a target list of 21 Be stars, of which six were actually observed. 
The resulting interference patterns are referred to as $S$-curves due to their shape resembling a sideways S. The FGS has a $5^"\times5^"$ instantaneous field of view which is then scanned by the instrument at a $45^\circ$ angle to its $x$- and $y$-axes. If the light source is from two stars, or an extended object, the appearance of the interference of the fringe pattern will differ from that  of a single point source i.e., a star \citep{InstHandbook}.  At any one time, two of the three sensors are used for pointing and guiding the telescope. This leaves one sensor available as a science instrument. The FGSs operate in two modes: POS or position mode; and TRANS or transfer mode. POS mode is mainly used for astrometry, with positional accuracies as high as $\sim1$ mas \citep{InstHandbook}. TRANS mode is used for high angular resolution observation, which we use in this investigation. Hubble FGS has higher precision even when compared to other Hubble instruments \citep{InstHandbook}. For comparison both Hubble/WFC3 and Hubble/ACS can only resolve down to $\sim200~$mas \citep{ACSInstHandbook}\\
\begin{figure}[ht!]
\centering
\includegraphics[width=0.5\textwidth]{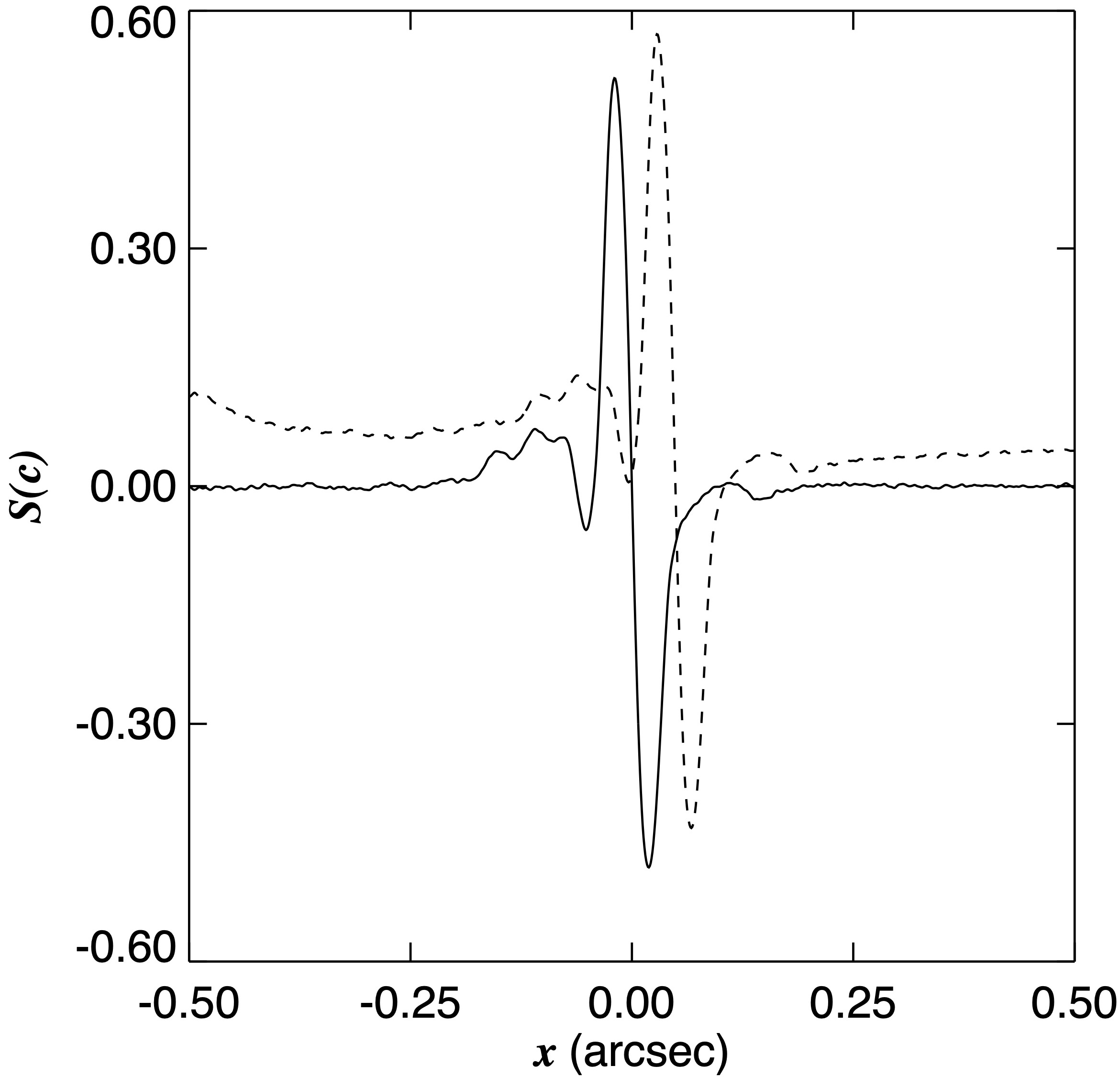}
\caption{Comparison of rectified (solid) and not rectified (dashed) $S$-curve of calibrator HD 50737. The $x$-axis is the instantaneous field of view of the FGS where 0 is the center of light.} \label{fig:Rectified}
\end{figure}\\

The data reduction process for $S$-curves is based on algorithms developed by the Space Telescope Science Institute (STScI) Astrometry team at Lowell Observatory \citep{Franz1991}. This includes correcting for drift and jitter of the spacecraft, co-adding the scans (15 for each target), and smoothing the final $S$-curve. All observations utilized the F583W filter, which has a central wavelength of 5830\AA \;and a band pass of 3400\AA. 
Individual scans that were either too noisy or required a large shift were rejected due to quality concerns. Finally, in order to perform a cross-correlation function (CCF) we need to rectify the $S$-curve by removing the variation from 0 due to the different sensitives in the the two photomultiplier tubes. We rectified the $S$-curve by subtracting a parabolic fit of the outer wings of the fringe. In cases where the parabolic fit did not represent the background well we used a multi-point spline fit. A rectified $S$-curve of HD 50737 is shown in Fig. \ref{fig:Rectified}. The dashed line is the observed $S$-curve and the solid black line is the rectified curve after the parabolic or spline fit.

\section{Companion Detection}
\label{sec:Comp Dect}\

$S$-curves, or interferograms, of binary stars display as a normalized superposition of two point sources,  shifted by their angular separation and scaled by their relative brightness. A TRANS Mode FGS observation produces two mutually orthogonal S-curves, referred to as the $x$ and $y$ components. Detection and measurement of a binary system is done by comparing models of binary system constructed from $S$-curves obtained from  an observation of a point source of similar B-V color (FGS optics includes refractive elements), with the constraint that the magnitude difference be held the same on both the $x$- and $y$-axis $S$-curves of binary systems that are not near the detection limit of the FGS or display clear departures from expected point source morphology. This is due to the optics of the setting of the instrument and color dependence of the appearance of the $S$-curves such as the profile widening for redder targets \citep{Horch2006}. If the difference between two presumably single-star $S$-curves from two targets is observed to be as expected from photometric noise, one can be confident both stars are unresolved and can be used as point source calibrators. This is due to the very unlikely probability that any two of the targets would have identical binary parameters (separation and differential magnitude). This process identified HD 52244 as being a potential binary.

Following the procedure set in \citet{Caballero2014}, we ran the STScI pipeline (hereafter STScI code) in order to fit a binary $S$-curve to HD 52244. The STScI code is a FORTRAN \& C code that fits a model using the least squares metric to fit a model $S$-curve to the observed one \citep{DataHandbook}. The model is created by superimposing two copies of a calibrator $S$-curve. The projected separation, $\rho$ and differential magnitude $\Delta\text{m}_{\text{F583W}}$ are seeded by the user and the algorithm varies these to minimize the least squares metric to fit a model curve and therefore determines the separation and differential magnitude of the system. In the case of the $x$-axis, we forced the algorithm to hold the magnitude difference from the easier fit $y$-axis. We selected HD 50737 as the calibrator since it was closest in color to HD 52244. We measured a separation of $\rho=42.2$ mas and a differential magnitude of $\Delta\text{m}_{\text{F583W}}$ = 1.91 mag.

In addition to the least squares fit, we applied a cross-correlation algorithm and performed a second-derivative test of the observed $S$-curve. These tests are sensitive to different separations and provided an additional method to support the detection of a component of small separation $\rho\leq 20$ mas. The cross-correlation is sensitive to when the projected separation is large ($\gtrsim 25\text{ mas}$). This test cross-correlates a calibrator $S$-curve with the non-single point source $S$-curve. This results in higher residuals of the cross-correlation function at the location of a companion. The flux ratio is then calculated based off of the scale factor for the two components.

\begin{figure*}[!ht]
\plotone{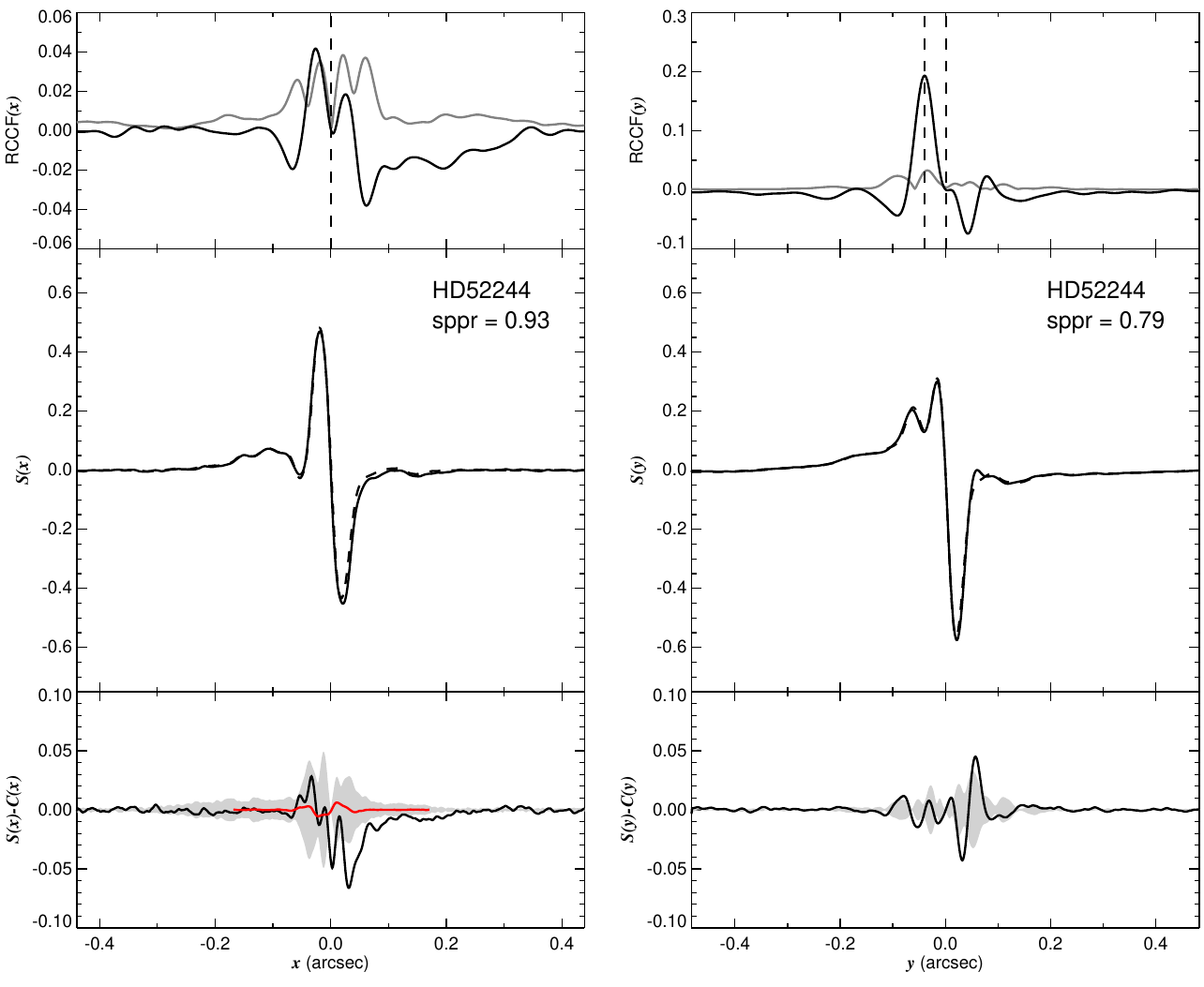}
\caption{This plot shows the model fit from the CCF and second derivative test for HD 52244 in $x$ (left) and $y$ (right). The top row is the mean residuals from the cross correlation routine. The solid black line is the CCF result after removing the primary component. The gray line is the calculated standard deviation and the vertical dashed line indicates where a component was detected (one in $x$, two in $y$). The middle row is the fit (dashed line) and the observed (solid line) from the CCF. The bottom row shows the results from the second derivative test. The solid black line is the observed minus model residuals and the gray region is the standard deviation of the ensemble of calibrator $S$-curves. The red line, present only on the $x$-axis plots, is the second derivative of the $S$-curve, indicating that while the system is not resolved in $x$ with the CCF, it deviates from a point source in $x$ with the second derivative test.\label{fig:HD52244_analysis}}
\end{figure*}

The second derivative test is sensitive to projected separations $\lesssim 25\text{ mas}$.  Since the observed $S$-curve for a binary star would be the superposition of two single $S$-curves we can mathematically describe the resulting $S$-curve as:
\begin{equation}
    S(x)_{obs}=\sum_{i=1}^nf_iS(x-x_i)
    \label{eqn:S-multi}
\end{equation}
where each $i$ star has a flux fraction $f_i=F_i/\Sigma F_j$ and a relative positional offset of $x_i$. For a system with only two components this becomes, 
\begin{equation}
    S(x)_{obs}=\frac{1}{1+r}S(x)+\frac{r}{1+r}S(x+\Delta x)
    \label{S-two}
\end{equation} where r is the flux ratio $r=\frac{F_2}{F_1}$
\citep{Caballero2014}.
By using a second order Taylor expansion we can see that the difference between a binary and single star (calibrator) is proportional to the second derivative of the calibrator,
\begin{equation}
    S(x)_{bin}-S(x)_{cal}=\frac{1}{2}\frac{r}{(1+r)^2}(\Delta x)^2S^"(x).
    \label{eqn:taylor}
\end{equation}
This result is further explained in \citet{Caballero2014}. Using Eqn. \ref{eqn:taylor}, we can compare the difference between the binary target and a calibrator (black line), with the second derivative of the calibrator (red line) in fig. \ref{fig:HD52244_analysis}. These two should be proportional with the constant of proportionality equaling $\frac{1}{2}\frac{r (\Delta x)^2}{(1+r)^2}$. In this form we are not able to determine the separation and flux ratio  without a secondary measurement of either.

The results of the cross-correlation and second derivative tests can be seen in the six panel plot in Figure \ref{fig:HD52244_analysis}, where the left and right sides are the $x$- and $y$-axes, respectively. The top row shows the results from the cross-correlation function with the vertical dashed line shows the location of detected components. The center main row is the observed $S$-curve in the solid black line with the constructed model based off the tests over-plotted with a dashed line. The bottom row shows the results of the second derivative test only in the $x$-axis indicating a companion in that axis not detected by the cross correlation function. These methods produce a fit with a separation of -41.0 mas in the $y$-axis as well as an unresolved detected companion in the $x$-axis. 

It should also be noted a disk around a star, as would be expected for a Be star, would also produce a non-point spread function $S$-curve. The extent of the disk for $\zeta\text{ Tauri}$, a star of similar spectral type, is thought to be 0.5 AU \citep{Rivinius2013}. HD 52244 has a distance as determined by Gaia of 1734 pc as well as hydrogen emission lines (specifically H$\alpha$) indicative of presently having a disk \citep{GaiaDR3} this would result in an angular extent of the disk of 0.88 mas this is well below the limit for FGS to resolve, meaning the deviation from a point-source $S$-curve is more likely due to a companion star.

\begin{figure*}
    \centering
    \plottwo{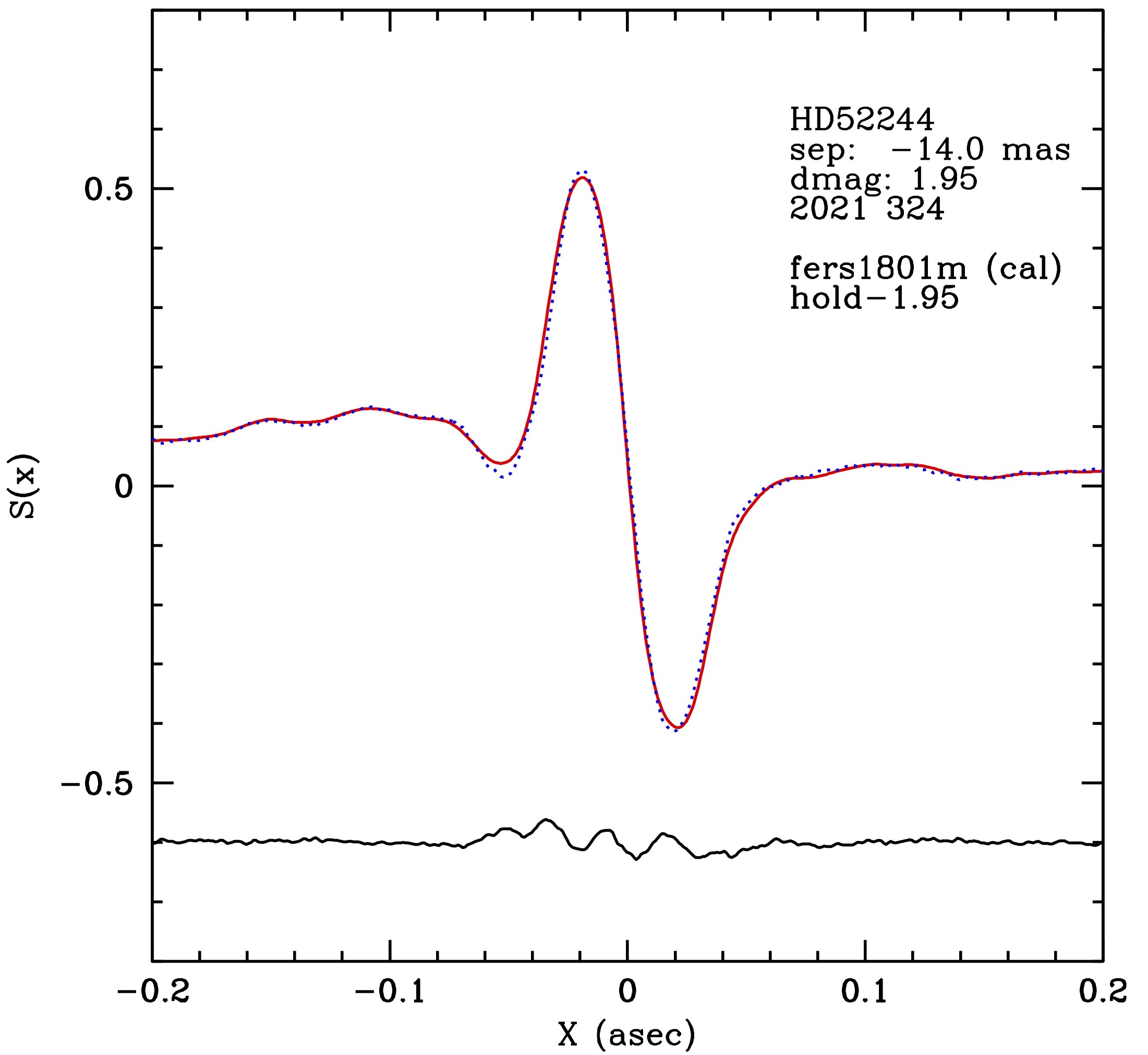}{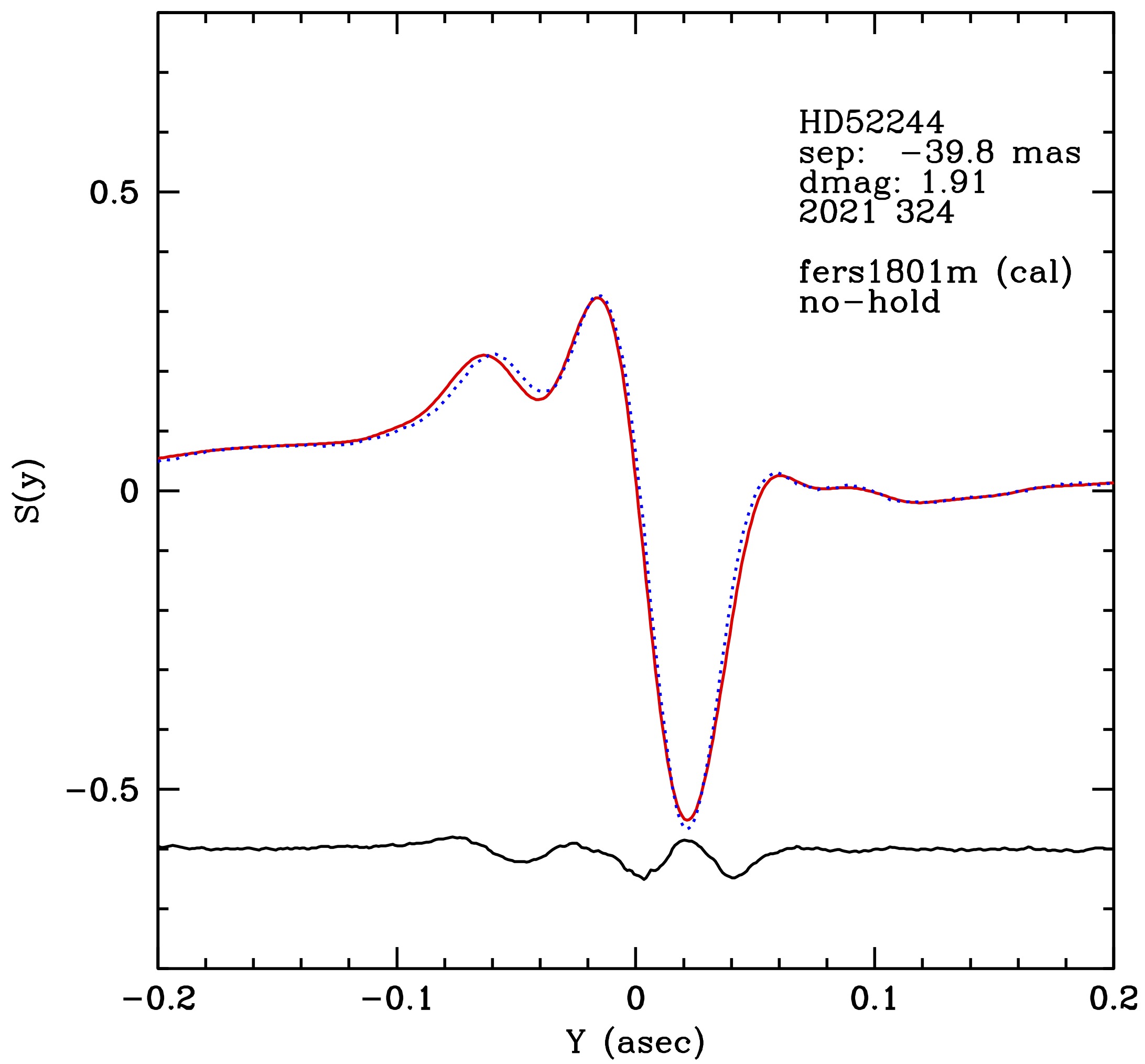}
    \caption{The $x$ (right) and $y$ (left) $S$-curves from the STSci least squares fitting for HD 52244. For both plots the red line is the observed $S$=curve and the blue dotted line is the model. The black line at the bottom is the difference between the two.}
    \label{fig:ObsVModel}
\end{figure*}

\section{Multiplicity of the Sample}
\label{sec:Discussion}
None of the stars in our sample have been previously identified as having companions. Three stars, HD 50737, HD 52244, and HD 206135 are identified as having a proper motion anomaly which may suggest a sub-stellar companion close to the primary star \citep{Kervella2019}. These companions would have separations and/of flux ratios too small to be detected in this investigation. \citet{Cruzalebes2019} observed HD 50737 and HD 206135 interferometrically using the Very Large Telescope Interferometer. In these cases, a binary companion would have been identified in the observations, however, they were able to angularly resolve the star. If a companion was present, this would have become evident during their reduction. HD 35345 was identified as presumably single the literature review by \citet{Bodensteiner2019}. \citet{Peter2012} observed HD 239618 using a lucky image technique and did not detect a companion. There are no citations indicating the binary status of HD 33232.

With the release of Gaia DR3 \citep{GaiaDR3} the re-normalized unit weight error (RUWE) value has been used as an indicator of possible binary status. The RUWE is related to the goodness of fit of the astrometric solution of the data. \citet{Castro2024} state that a RUWE value of greater than 1.4 indicates the possible presence of a companion. Be stars being a star with a disk adds complications to the astrometric fit of the Gaia data and inflates the RUWE value \citep{Fitton2022}. Due to this artificial inflation, the nominal value of 1.4 cannot be used for classification. In the sample used by \citet{Fitton2022}, they use NEOWISE photometry (W1-W3 $>1$) to identify stars with a disk. They adopt a value of RUWE $>$ 2.5 as the threshold for stars containing a disk and likely to also show a binary anomaly in their Gaia measurements. The issue arises with Be stars whose disks are temporally varying and may even cease to exist at some points. For this reason, we looked at NEOWISE multi-epoch photometry to determine if at any time during the Gaia DR3 observations the W1-W3 color surpassed 1 for our target stars, suggesting the need to use the RUWE$>$2.5 criterion for potential binarity. All stars have a color greater than 1 and a RUWE value less than 2.5, with the exception of HD 206135, whose color was less than 1 but has a RUWE less than 1.4. Following the criteria set forth in \citet{Fitton2022}, none of these stars surpass the RUWE value indicative of a potential companion based off of the NEOWISE W1-W3 color, including HD 52244, which we resolved a companion for using FGS (see discussion below) . 

\subsection{Discussion of HD 52244 as a new binary detection}
Of the six stars observed, only one star HD 52244, displayed signs of binarity. Results from both the STScI code from \citet{DataHandbook} as well as the CCF and second derivative tests from \citet{Caballero2014} are in agreement that the separation on the $y$-axis is $\sim-40.0\text{ mas}$ and differential magnitude on the $y$-axis  of the system is $\sim1.91$ mag. The second-derivative routine does not get a numerical separation for $x$-axis since the separation and differential magnitude are entangled in the coefficient of the second derivative from the Taylor expansion (as described in section 3). The second derivative test does identify a companion present since the shaded grey region (standard deviation) in the bottom left panel of Fig. \ref{fig:HD52244_analysis} is larger than the residuals. The separation and flux ratio with this method cannot be disentangled without a secondary solution of the flux ratio, using the flux ratio in the $x$-axis from the STScI code produces a result that is  beyond the ability of the telescope and therefore deemed unrealistic. The STScI code does determine that the separation in the $x$-axis is $-14.0\text{ mas}$. With all these pieces we can determine the total separation and position angle of the system to be $144^\circ$  East of North and a total separation of $42.2\text{ mas}$. These values are consistent across all methods used. HD 52244 is not currently known to have any companion, visual or otherwise, (other than the possible proper motion anomaly discussed in Sec. \ref{sec:Sample} whose separation and differential magnitude are vastly different from our detection) meaning this is a new detection of a companion for this star.

In order to provide preliminary characterization of the system, we adopt $\Delta m_V =1.91$, based on the differential magnitude in the F583W filter, keeping in mind that F583W is a much broader filter than Johnson-Cousin V. If we assume that the Be star is on the main sequence, then according to \citet{Pecaut&Mamajek2013} we approximate the companion to be an early B8~V star with $M_V=0.11$. This gives us a total mass of the system of $M_1(7.3\:M_\odot)+ M_2(3.38\:M_\odot)= 10.68\:M_\odot$. Assuming a circular orbit, and that the system is face-on, a projected separation of 42.7 mas corresponds to a  radius of 74 AU. Using Kepler's $3^{rd}$ , we estimate this system would have a period of about 110 years. Follow-up observations with FGS should be able to detect such motion.

\begin{deluxetable*}{CCC}[ht!] %use capitals to Keep columns in mathmode
\tablewidth{0pt}
\tablecaption{Multiplicity parameters for HD 52244 with the STSci Code and CCF \& Second Derivative tests.}
\tablewidth{0pt}
\tablehead{
\nocolhead{}&\colhead{STScI}&\colhead{CCF}
}
\startdata
\Delta x&-13.8 \pm 0.2 \text{ mas} &\nodata\\
\Delta y& -39.8 \pm 0.2 \text{ mas}&-41.0\text{ mas}\\
\rho& 42.2 \text{  mas}& 43.3\text{ mas}\\
& 73.3 \text{  AU}& 75.1\text{ AU}\\
\theta& 144.5^\circ&143.9^{\circ,b}\\
\Delta \text{m}_x& 1.91&\nodata\\
\Delta \text{m}_y&1.91 &1.90\\
\Delta \text{m}&1.91^a &\nodata
\enddata
\tablecomments{$^a\Delta \text{m}$ This is fit in the $y$ axis and then held for the $x$-axis. $^b$ These results use the  $\Delta x$ from the STSci code since it was not resolved with the CCF or second derivative test.}
\end{deluxetable*}
\section{Summary \& Future Work}
\label{sec:sum+FW}
%B stars are the progenitors to such events as supernovae (SNe) and Cepheid Variables. Characterizing the multiplicity of B stars is essential for predicting their class of core collapse SNe.
In this paper we present the results of an HST/FGS survey of six Be stars, maximizing the use of the instrument during downtime due to the Science Instrument Command and Data Handling module failure in November of 2021. Only one of these stars, HD 52244, displays an $S$-curve suggesting a non-point source (Figure \ref{fig:ObsVModel}, right panel). This is a new detection of a companion for HD 52244. For the remaining five targets we do not detect a companion with a separation larger than 10 milli-arcseconds (9-13 AU). Using two different approaches, we determined the separation of HD 52244 to be $\rho=42.8\pm0.6$ mas with a position angle of $\theta=144.4^\circ\pm0.1$ and a differential magnitude of $\Delta m_{583W}=1.91\pm0.01$ in the HST F583W filter.  \citet{Kervella2019} does show that this star has a proper motion anomaly that may suggest a companion, however, their predicted mass to distance ratio is either below the distance limit or below the brightness at the distances we are sensitive to.
This work establishes that HST/FGS is a powerful instrument to probe the binary status of Be stars including detections of wide third companions to these that may be the key to understating result of angular momentum being placed into that third companion as described in \citet{Tokovinin1997}. We hope to combine with work with other surveys, such as Gemini Speckle survey \citep{Kalari2025}, and with future works (e.g. Kamp in prep) employing high angular resolution methods to further constrain the multiplicity fraction of this interesting class of stars.
\section*{Acknowledgments}
This research has made use of the Be Star Spectra (BeSS) database \citep{Neiner2018}.
This research has made use of the SIMBAD database, operated at
CDS, Strasbourg, France \citep{Wegner2000}.
This research was made possible by STSci grant HST-GO-16868.007-A, the Ortega Fellowship, and Dr. Quiroga-Nu\~nez at Florida Tech.
\bibliography{Cites}
\bibliographystyle{aasjournal}
\appendix
\section{Plots from IDL}
The following plots are from the IDL CCF and second derivative algorithms for each of the other observed stars.
\begin{figure}
\includegraphics[width=\textwidth]{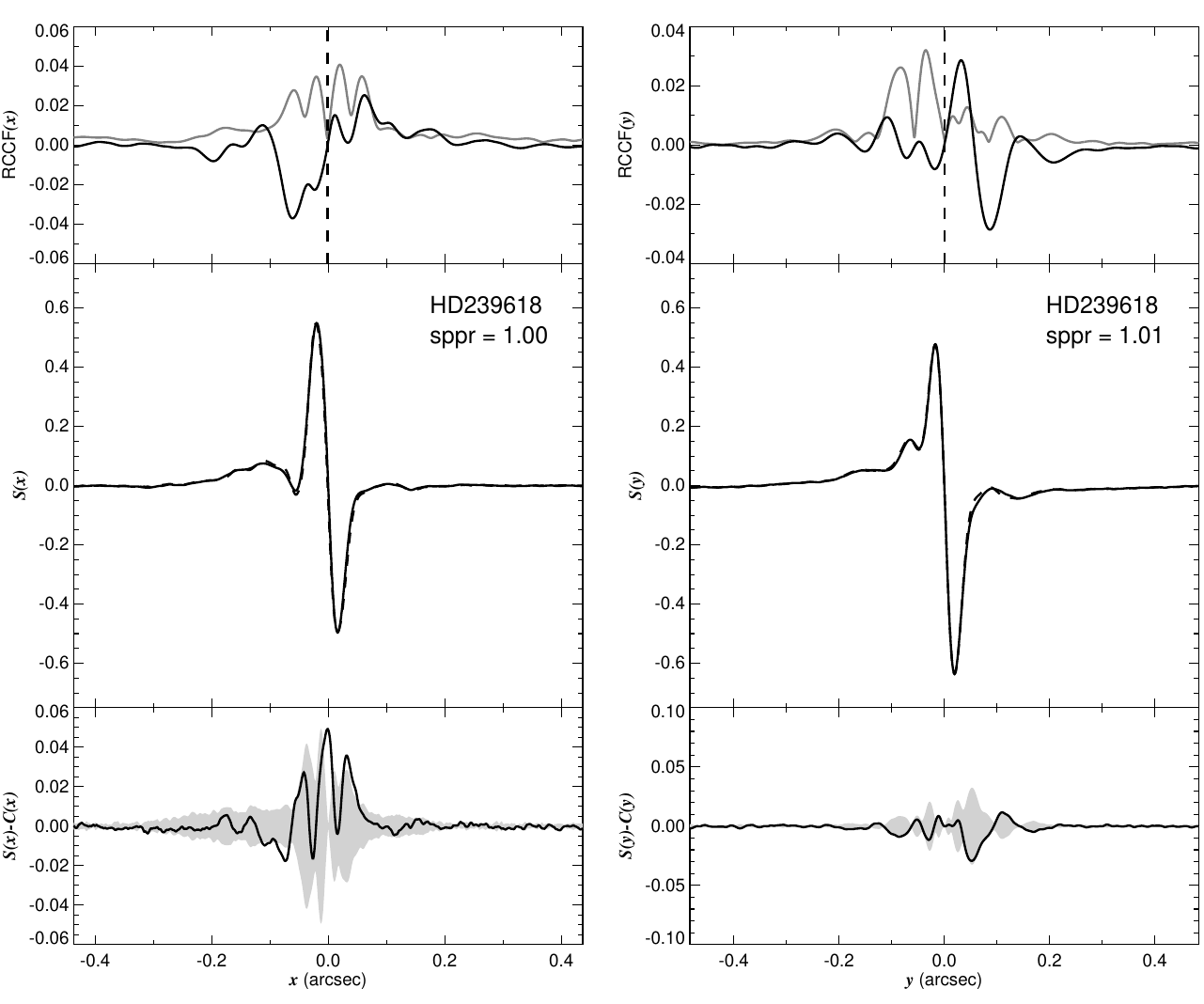}
\caption{This plot shows the model fit from the CCF and second derivative test for HD 239618 in $x$ (left) and $y$ (right). The top row is the mean residuals from the cross correlation routine. The solid black line is the CCF result after removing the primary component. The gray line is the calculated standard deviation and the vertical dashed line indicates where a component was detected. The middle row is the fit (dashed line) and the observed (solid line) from the CCF. The bottom row shows the results from the second derivative test. The solid black line is the observed minus model residuals and the gray region is the standard deviation of the ensemble of calibrator $S$-curves.}
\end{figure}
\newpage
\begin{figure}
\includegraphics[width=\textwidth]{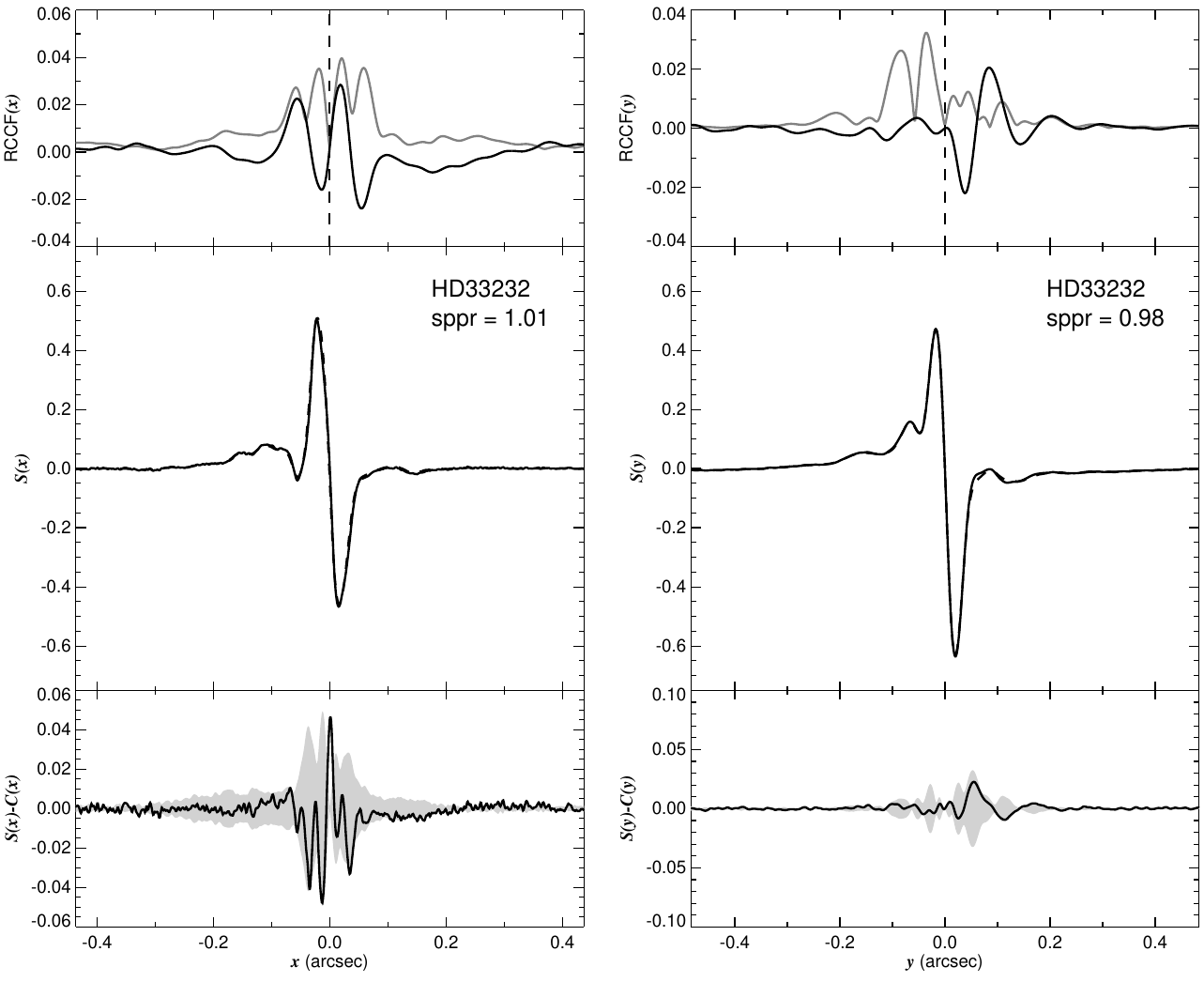}
\caption{This plot shows the model fit from the CCF and second derivative test for HD 33232 in $x$ (left) and $y$ (right). The top row is the mean residuals from the cross correlation routine. The solid black line is the CCF result after removing the primary component. The gray line is the calculated standard deviation and the vertical dashed line indicates where a component was detected. The middle row is the fit (dashed line) and the observed (solid line) from the CCF. The bottom row shows the results from the second derivative test. The solid black line is the observed minus model residuals and the gray region is the standard deviation of the ensemble of calibrator $S$-curves.}
\end{figure}
\newpage
\begin{figure}
\includegraphics[width=\textwidth]{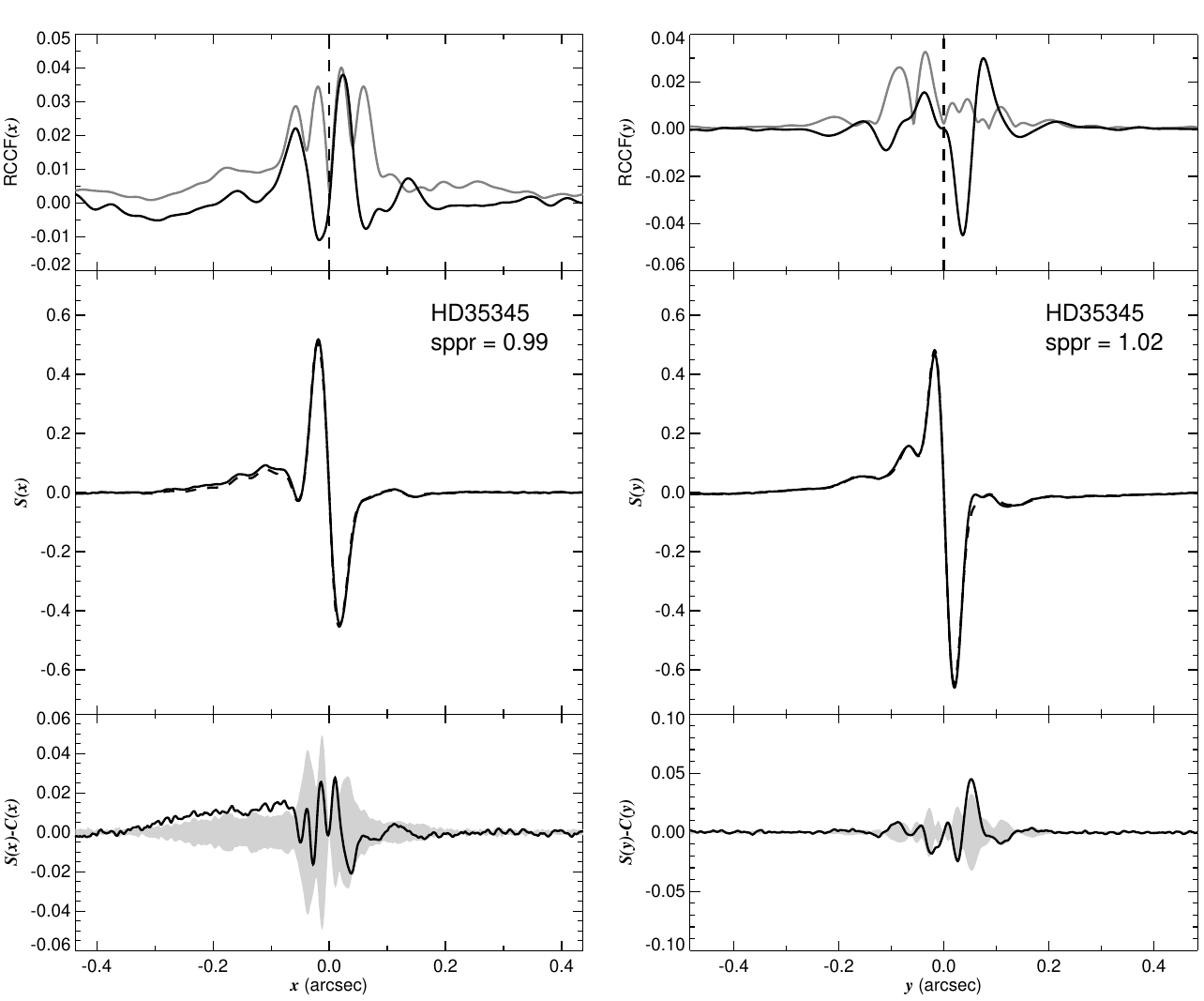}
\caption{This plot shows the model fit from the CCF and second derivative test for HD 35345 in $x$ (left) and $y$ (right). The top row is the mean residuals from the cross correlation routine. The solid black line is the CCF result after removing the primary component. The gray line is the calculated standard deviation and the vertical dashed line indicates where a component was detected. The middle row is the fit (dashed line) and the observed (solid line) from the CCF. The bottom row shows the results from the second derivative test. The solid black line is the observed minus model residuals and the gray region is the standard deviation of the ensemble of calibrator $S$-curves.}
\end{figure}
\newpage
\begin{figure}
\includegraphics[width=\textwidth]{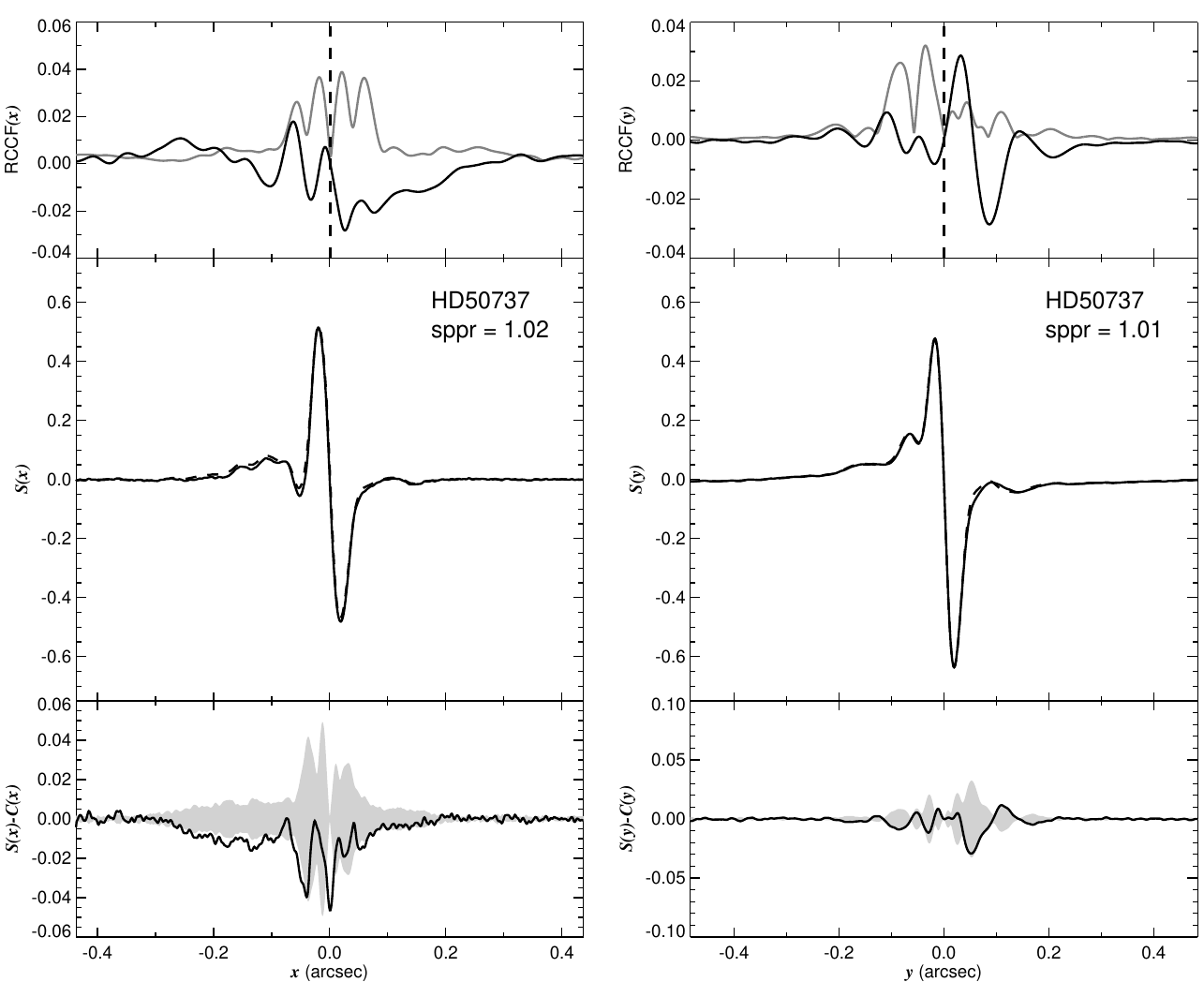}
\caption{This plot shows the model fit from the CCF and second derivative test for HD 50737 in $x$ (left) and $y$ (right). The top row is the mean residuals from the cross correlation routine. The solid black line is the CCF result after removing the primary component. The gray line is the calculated standard deviation and the vertical dashed line indicates where a component was detected. The middle row is the fit (dashed line) and the observed (solid line) from the CCF. The bottom row shows the results from the second derivative test. The solid black line is the observed minus model residuals and the gray region is the standard deviation of the ensemble of calibrator $S$-curves.}
\end{figure}
\begin{figure}
\newpage
\includegraphics[width=\textwidth]{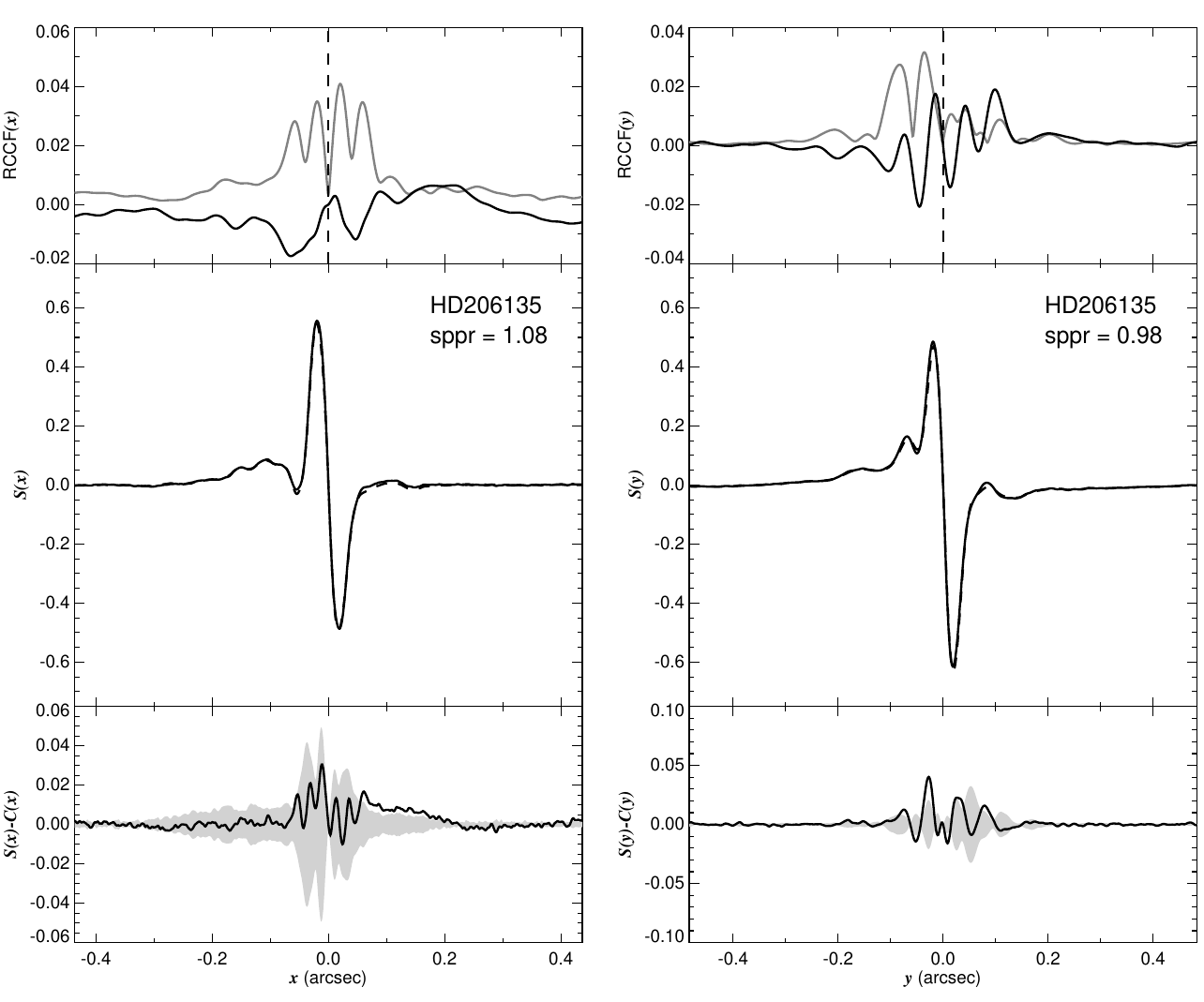}
\caption{This plot shows the model fit from the CCF and second derivative test for 206135 in $x$ (left) and $y$ (right). The top row is the mean residuals from the cross correlation routine. The solid black line is the CCF result after removing the primary component. The gray line is the calculated standard deviation and the vertical dashed line indicates where a component was detected. The middle row is the fit (dashed line) and the observed (solid line) from the CCF. The bottom row shows the results from the second derivative test. The solid black line is the observed minus model residuals and the gray region is the standard deviation of the ensemble of calibrator $S$-curves.}
\end{figure}
\end{document}